\documentclass[fleqn,12pt,twoside]{article}
\usepackage{espcrc1}

\usepackage{graphicx}
\usepackage[figuresright]{rotating}
 
\newcommand{\AmS}{{\protect\the\textfont2
  A\kern-.1667em\lower.5ex\hbox{M}\kern-.125emS}}

\title{Kinematical contributions to the transverse asymmetry \\ in
semi-inclusive DIS}

\author{K. A. Oganessyan\address[LNF]{INFN-Laboratori Nazionali di Frascati I-00044 Frascati, \\ 
        via Enrico Fermi 40, Italy} \address[DESY]{DESY, Notkestrasse 85, 
        22603 Hamburg, Germany}, 
        P. J. Mulders\address{Division of Physics and Astronomy, Vrije Universiteit 
        De Boelelaan 1081, \\
        NL-1081 HV Amsterdam, the Netherlands}, 
        E. De Sanctis\addressmark[LNF],
        and
        L. S. Asilyan\addressmark[LNF]}
       
\begin{document}

\maketitle

\begin{abstract}
We discuss the contributions of the transverse spin component 
of the target to the double-spin asymmetries in semi-inclusive 
deep inelastic scattering of longitudinally polarized electrons 
off longitudinally polarized protons.     
\end{abstract}

\vspace*{1.0cm}

In the studies of semi-inclusive charged and neutral pion production off a 
longitudinally polarized protons,  the HERMES collaboration has observed  
a single target-spin asymmetry (SSA)~\cite{HERM}. This asymmetry could 
either result from twist-3 chiral-odd effects~\cite{TM,AK} and/or could 
be a reflection of the Collins effect~\cite{COL}. Which of the two is 
relevant is an open issue. For a further understanding of ``transverse 
asymmetry contribution'' we discuss here the double-spin and double-spin 
azimuthal asymmetries, where those contributions are well defined.      

A target with an anti-parallel 
(parallel) polarization with respect to the beam has a transverse spin component 
in the virtual photon frame which can only have  azimuthal angle $\pi$ ($0$) (Fig.1). 
The value of this transverse spin component is  
\begin{equation}
\label{ST}
 \vert S_T \vert = \vert S \vert \sin{\theta}_{\gamma}, 
\end{equation}
where $S$ is target polarization.
The quantity $\sin\theta_{\gamma}$ is of
order $1/Q$ and is given by
\begin{equation}
\label{ST}
\sin\theta_{\gamma} = 
\sqrt{\frac{4M^2x^2}{Q^2+4M^2x^2} (1-y-{M^2x^2 y^2 \over Q^2})},
\end{equation}
where $M$ is the nucleon mass. 

\includegraphics[width=16pc,height=22pc]{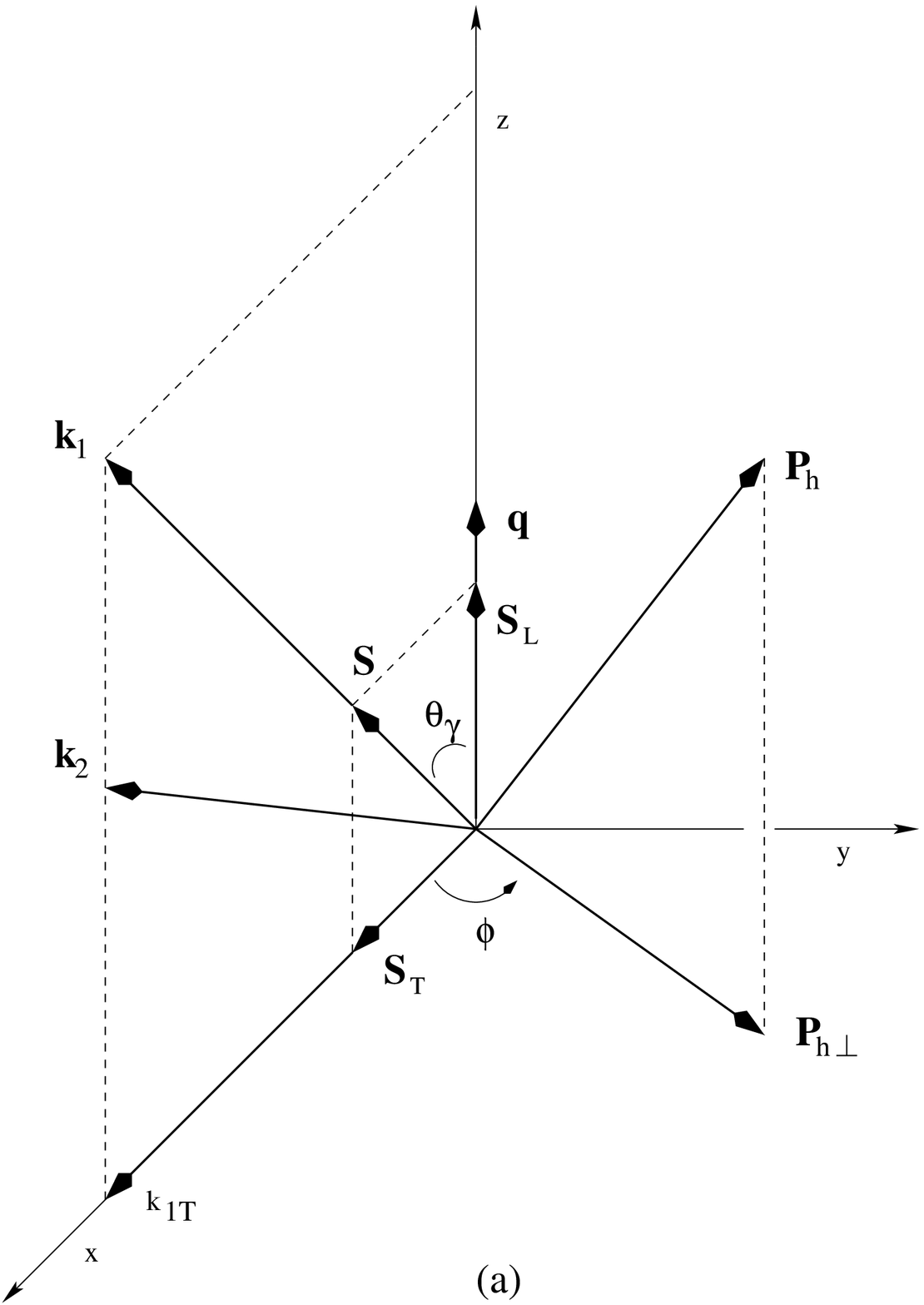}~~~~~~~~~~ 
\includegraphics[width=15pc,height=20pc]{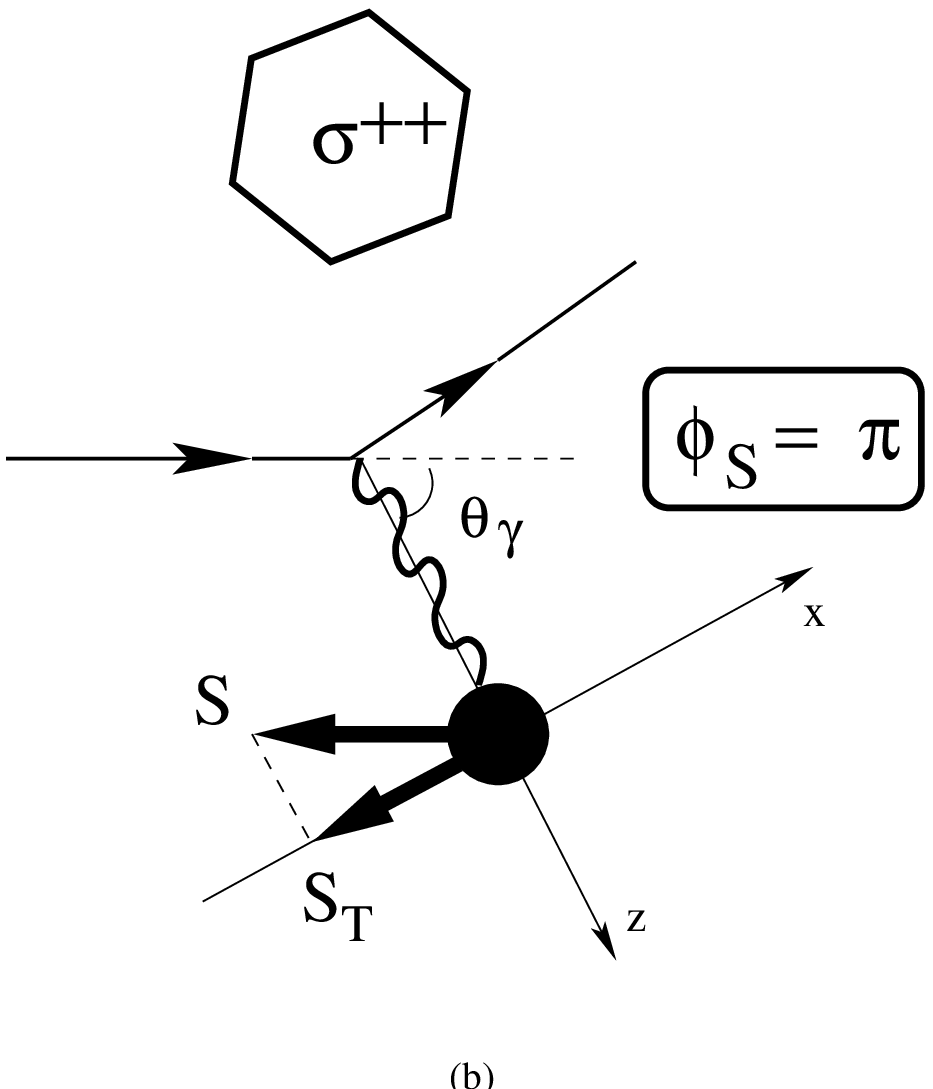}\\ \\ 
Fig.1. (a) -- The kinematics of semi-inclusive DIS: $k_1$ ($k_2$) is 
the 4-momentum of the 
incoming (outgoing) charged lepton, $Q^2=-q^2$, where $q=k_1-k_2$, 
is the 4-momentum of the virtual photon. The momentum
$P$ ($P_h$) is the momentum of the target (observed) hadron. The scaling
variables are $x=Q^2/2(P\cdot q)$, 
$y=(P\cdot q)/(P\cdot k_1)$, and $z=(P\cdot P_h)/(P\cdot q)$. 
The momentum $k_{1T}$ ($P_{h\perp}$) is the incoming lepton (observed hadron) 
momentum component perpendicular to the virtual photon momentum direction, 
and $\phi$ is the azimuthal angle between $P_{h\perp}$ and $k_{1T}$. 
(b) -- The definition of the azimuthal angle $\phi_S$ and the target
polarization components in virtual photon frame. 

\vspace{0.4cm}

First, we give an estimate of 
the $\cos\phi$ moment of the semi-inclusive DIS cross 
section, which is the following
weighted integral of a cross section asymmetry~\cite{OMD},
\begin{equation}
A^{\cos\phi}_{{LL}} = {1 \over {\langle P_{h\perp} \rangle} } 
\frac{\int d^2P_{h\perp} {\vert P_{h\perp}\vert}
\cos \phi \left(\sigma^{++}+\sigma^{--}-\sigma^{+-}-\sigma^{-+} \right)}
{{1 \over 4}\int d^2P_{h\perp} \left(\sigma^{++}+\sigma^{--}+
\sigma^{+-}+\sigma^{-+} \right)}.
\label{ASMY}
\end{equation}
Here the subscript $LL$ denotes the longitudinal polarization of the beam and target 
respectively, 
$\sigma$ is a shorthand notation for  
${d \sigma^{eN \to e h X} /{dx\,dy\,dz\,d^2P_{h\,\perp}}}$, the superscripts 
$++,-- (+-,-+)$ denote the helicity states of the beam and target respectively, 
corresponding to antiparallel (parallel) polarization\footnote{It leads to positive 
$g_1(x)$. }. Assuming $100 \%$ beam 
and target polarization and using the Wandzura-Wilczek (WW) approximation~\cite{WW}, 
where only the twist-2 distribution 
and fragmentation functions are used, i.e. the interaction-dependent 
twist-3 parts are set to zero, one obtains (for more details see Ref.~\cite{OMD})   
\begin{equation}
A^{\cos\phi}_{LL} = {4 \over {\langle P_{h\perp} \rangle} } 
\,\frac{\Delta\sigma_{LL} - d \sigma_{LT}}{\sigma_{UU}},   
\label{AS}
\end{equation}
where 
\begin{equation}
\Delta\sigma_{LL} \stackrel{\mathrm{WW}}{\approx} 
- 4 \lambda_e {S_L \over Q}\, \sqrt{1-y}\,   M^2 g_1^{(1)}(x) z D_1(z),
\label{AS1}
\end{equation}
\begin{equation}
 d\sigma_{LT} \stackrel{\mathrm{WW}}{\approx} 
\lambda_e \vert S_T \vert \, (2-y)\, M\,  
\left[\int_x^1 du\ \frac{g_1(u)}{u}\right]
z\, D_1(z),
\label{AS2}
\end{equation}
\begin{equation}
\sigma_{UU} = \frac{[1+(1-y)^2]}{y}\,f_1(x) D_1(z), 
\label{AS3}
\end{equation}
being $f_1$ and $g_1$ ($D_1$) the well-known leading twist distribution (fragmentation) 
functions. Notice that the cross section $d\sigma_{LT}$ is positive but gives a negative
contribution to the asymmetry (\ref{AS}) because of the dependence on the azimuthal angle 
$\phi_S$: 
$\sigma^{++}(\sigma^{--}) = - d\sigma_{LT}$ and $\sigma^{-+}(\sigma^{+-}) 
= d\sigma_{LT}$ at $\phi_S=\pi(0)$ (see Fig.1 (b)).

It is important to point out that in the WW approximation the $\cos\phi$
asymmetry reduces to a kinematical effect conditioned by intrinsic
transverse momentum of partons similar to the $\cos\phi$ asymmetry in
unpolarized semi-inclusive DIS~\cite{CAHN}. 

\includegraphics[width=18pc,height=18pc]{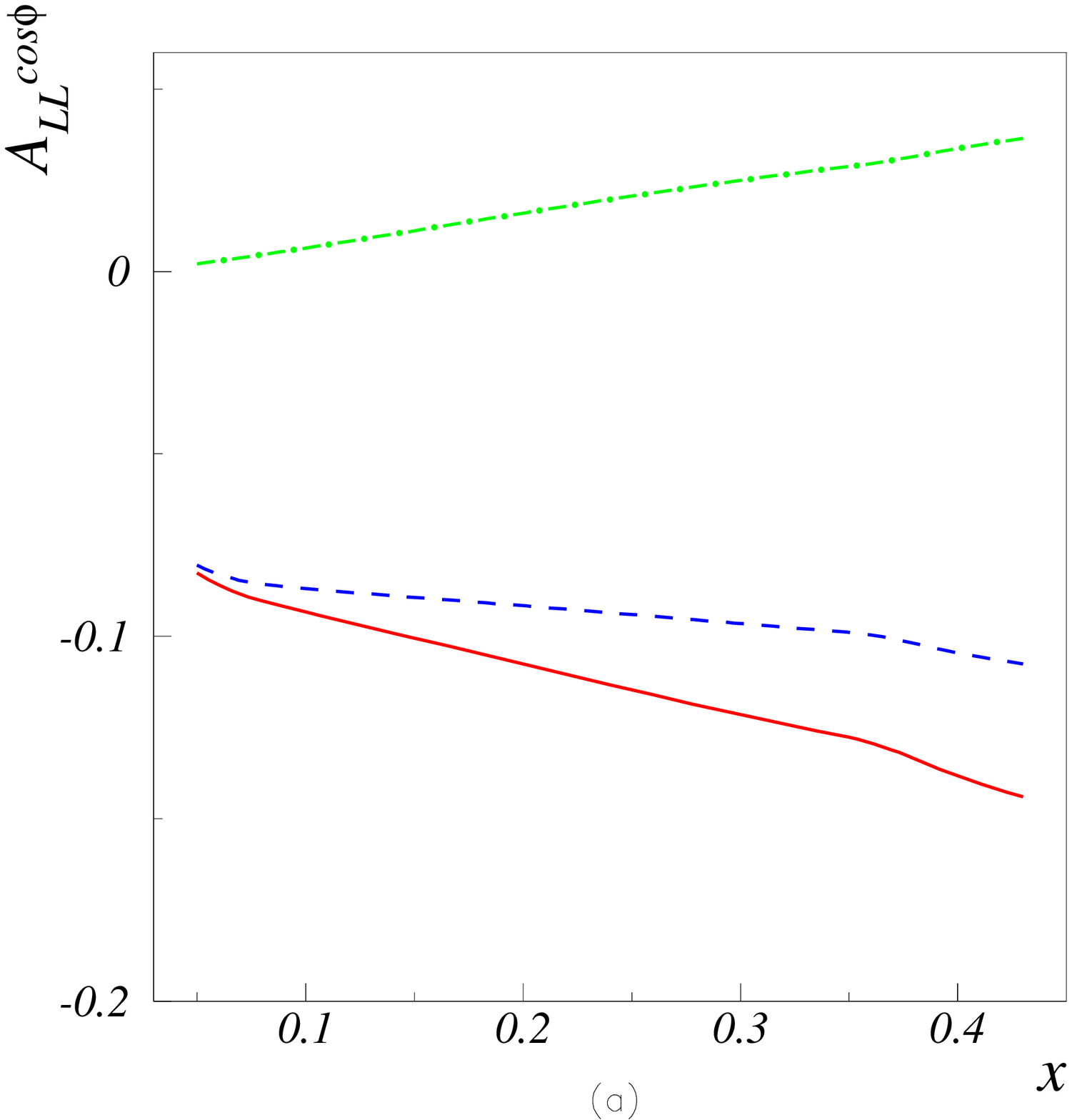}~~~~~~~~~~
\includegraphics[width=18pc,height=18pc]{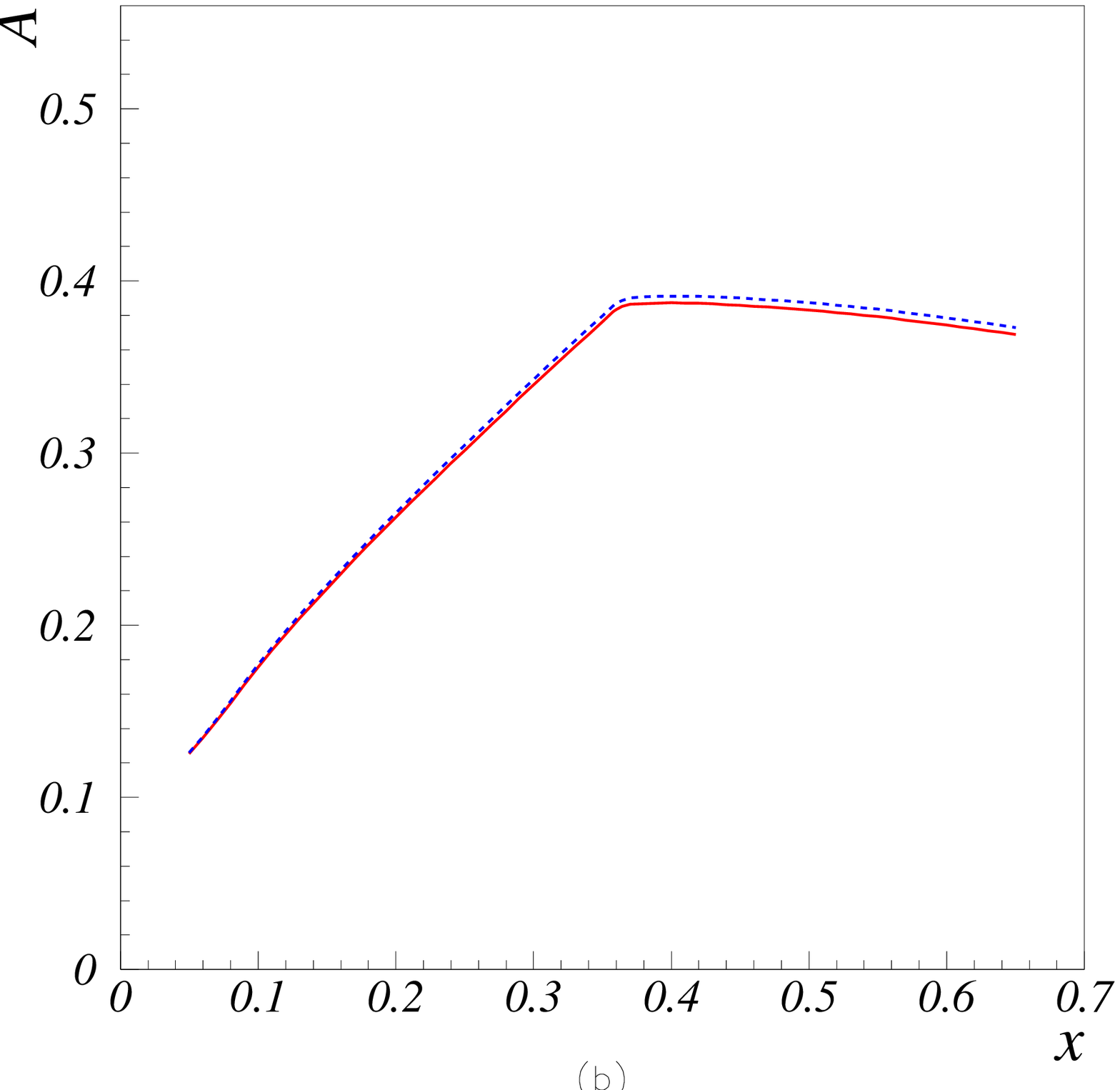}\\ \\
Fig.2. (a) -- $A_{LL}^{\cos\phi}$ of Eq.(\ref{AS}) for $\pi^{+}$
production as a function of $x$. The dashed line corresponds to 
contribution of the $\Delta \sigma_{LL}$, dot-dashed one to 
$d \sigma_{LT}$ and the solid line is the difference of those two; 
(b) -- Double-spin asymmetry, defined by Eq.(\ref{ASY1}), as a function of $x$. 
The full-curve corresponds to $\Delta \sigma^{'}_{LL}$ contribution and the dashed 
one is the total asymmetry.  

\vspace*{0.4cm}

In Fig.2(a), the asymmetry $A_{LL}^{\cos\phi}$ for $\pi^+$ production 
on a proton is shown as a function of $x$. The curves are calculated by 
integrating over the HERMES kinematical ranges~\cite{OMD}. As it can be seen, 
the WW approximation gives the large negative 
double-spin $\cos\phi$ asymmetry; the ''kinematic'' contribution coming from 
the transverse component of the target polarization is small (up to $25 \%$ 
at large $x$). 

Let us now consider the following asymmetry 
\begin{equation}
A =  
\frac{\int d^2P_{h\perp} 
\left(\sigma^{++}-\sigma^{-+} \right)}
{\int d^2P_{h\perp} \left(\sigma^{++}+\sigma^{-+} \right)},
\label{ASY1}
\end{equation}
which can be written as~\cite{TM,JJ}
\begin{equation}
A = \frac{\Delta \sigma^{'}_{LL} + d \sigma^{'}_{LT}}
{\sigma_{UU}}, 
\label{CSS}
\end{equation}
with
\begin{equation}
\Delta \sigma^{'}_{LL} = \lambda_e S_L\,(2-y)\,g_1(x)D_1(z), 
\label{CS1}
\end{equation}
\begin{equation}
d\sigma^{'}_{LT} \stackrel{\mathrm{WW}}{\approx} 4 {M \over Q} \vert S \vert 
y \sqrt{1-y} \, x^2 [ \int_x^1 du {g_1(u) \over u} ] \, D_1(z). 
\label{CS2}
\end{equation} 

\vspace*{0.4cm} 

In Fig.2(b), this asymmetry is given as a function of $x$. As it is shown, the 
contribution from the target transverse component is negligible.  

Another possibility for studying the ``kinematical'' contributions is 
considering 
of $\sin(2\phi-\phi_S)$ -- weighted asymmetry, $A^{\sin(2\phi-\phi_S)}_{LT}$, 
and its contribution to the target longitudinally polarized case.

\vspace*{0.3cm}

In summary, the double-spin and the double-spin azimuthal asymmetries of 
semi-inclusive DIS of 
longitudinally polarized electrons off longitudinally polarized protons 
at twist-two level was investigated. A sizable negative $\cos\phi$ asymmetry 
is found for HERMES kinematics; the `kinematical' contribution 
from target transverse component ($S_{T}$) to the $\cos\phi$ asymmetry, 
$A_{LL}^{\cos\phi}$, is small and that to the double-spin asymmetry, $A$, is 
negligible. Then, the measurements of SSA with transversely polarized 
target could help to understand the transverse asymmetry effects in the 
longitudinally polarized target case. 

This work is part of the research performed under the European Commission 
IHP program under contract HPRN-CT-2000-00130.

\end{document}